# Adaptive Template Enhancement for Improved Person Recognition using Small Datasets

Su Yang, Sanaul Hoque and Farzin Deravi, *Member*, IEEE

*Abstract* — **A novel instance-based method for the classification of electroencephalography (EEG) signals is presented and evaluated in this paper. The non-stationary nature of the EEG signals, coupled with the demanding task of pattern recognition with limited training data as well as the potentially noisy signal acquisition conditions, have motivated the work reported in this study. The proposed adaptive template enhancement mechanism transforms the feature-level instances by treating each feature dimension separately, hence resulting in improved class separation and better query-class matching. The proposed new instance-based learning algorithm is compared with a few related algorithms in a number of scenarios. A clinical grade 64-electrode EEG database, as well as a low-quality (high-noise level) EEG database obtained with a low-cost system using a single dry sensor have been used for evaluations in biometric person recognition. The proposed approach demonstrates significantly improved classification accuracy in both identification and verification scenarios. In particular, this new method is seen to provide a good classification performance for noisy EEG data, indicating its potential suitability for a wide range of applications.**

## I. INTRODUCTION

WITH the recent developments in the field of automatic pattern recognition, particularly the thriving area of deep learning techniques, pattern classification for non-stationary time-series modalities (such as natural language processing and speech recognition) have achieved considerable improvements [1][2]. However, pattern recognition for complex neurological signals such as electroencephalography (EEG) remains a challenging topic, and has been receiving an increasing attention in the research community [3]. In this paper, we propose a novel instance-based classification method for EEG pattern recognition in biometrics applications. To demonstrate its generality, the method is firstly evaluated using a public benchmarking dataset. EEG classification with a limited amount of training data has been also employed to further explore the effectiveness of the proposed algorithm in dealing with challenging pattern recognition problems.

Compared to eager learning algorithms which perform explicit generalization [4] (e.g., neural networks), one advantage of instance-based learning is its ability to flexibly model the training set by using only a subset of it, hence potentially provide better local approximations. As one important branch of machine learning, instance-based learning is designed to compare the query instance(s) with training instances, and store the results of these comparisons for subsequent decision-making [5]. For example, in a scenario where the size of the training data is dynamic (e.g., additional data enrolment), the templates of interest could be updated adaptably when using instance-based methods, the previously optimized performance may then be well maintained. However, increasing the number of templates without a selection scheme also introduces complexity to the system. Various instance reduction and generation algorithms have been developed to deal with such shortcomings of this type of algorithms [6][7]. The development of these template (prototype) optimization algorithms has had a number of aims, including 1) speed increase, 2) storage minimisation, 3) improving the learning speed and 4) improved generalization accuracy [6].

One of the earliest instance reduction algorithms is based on a selection rule namely condensed nearest neighbour (CNN) [8]. The traditional CNN, however, may not be able to correctly classify some of the instances due to the equal-weighted evaluation in the feature space [9], given the center(s) of the feature cluster(s) is(are) often considered to be heavily weighted than the outliers in classification. Efforts have been made since then to address and improve various aspects of the instance-based algorithms [6][10][11]. Typical instance-based learning algorithms where instance selection is paramount include the k-Nearest Neighbour (k-NN) and the Support Vector Machine (SVM) algorithms [9].

Limited availability of training data has long been a challenge for many pattern recognition problems. This may lead to under-generalization of the feature space and as such, result in poor classification performance in real-world scenarios where data volume is much larger than in the experimental scenario. To alleviate this, the algorithm proposed in this work aims to improve the pattern recognition through a multi-stage adaptive mechanism, incorporating by generating new templates through the reconstruction of the existing template data.

The within-class and between-class similarities are critical factors for effective classifications [12]. High between-class similarity may result in a high false positive rate while low within-class similarity may produce a high false negative classification rate. To achieve good recognition performance, instance distributions may be examined to isolate and select key

Su Yang is with the Department of Computer Science, Faculty of Science and Engineering, Swansea University, Swansea, UK (e-mail: su.yang@swansea.ac.uk)
Sanaul Hoque and Farzin Deravi are with the School of Engineering, University of Kent, Canterbury, UK.

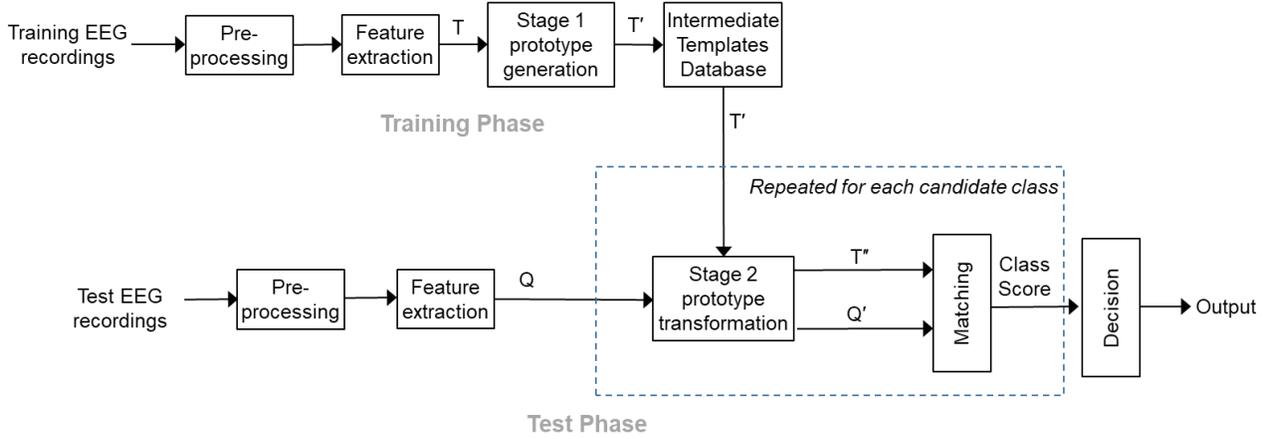

Fig. 1. The proposed I-ATR approach; T and Q indicate the training set and test set, T' and T'' are the intermediate templates generated with optimised between-class separation and within-class similarity.

instances with lower between-class similarity and higher within-class similarity. Such subsets of instances can then be used as building blocks to further remodel and provide compact representations for each class. This idea is further explored and implemented using the proposed technique, by subjecting the reconstruction process to be the mutually influenced by the query instance and the training templates, which are simultaneously remodeled for every classification attempt.

Certain time-series data such as EEG signals are characterized by their non-stationary behavior [12][13]: the signal statistics for the same class is often unstable and varies over time. The within-class variance over time, such as longitudinal template ageing effects in person recognition systems, is a great impacting factor in EEG-based pattern recognition. Based on our experimental investigations and other related reports, even an interval of several minutes between independent recordings could cause considerable template variations [3]. Therefore, data-driven optimum instance generation may play an important role for concurrently achieving both satisfactory within-class as well as between-class discriminations. Therefore, to explore the full potential of the proposed method, EEG signals, with their time-series nature, provide a suitable test case. This is the main motive of choosing EEG data for the algorithm evaluation in this work. Two distinctively different EEG databases are used to evaluate the effectiveness of the proposed instance-based template reconstruction learning algorithm.

The paper is structured as follows: The detailed description of the algorithm is given in Section II. The performance of the proposed method is evaluated in Section III, including a number of pattern recognition scenarios using EEG, along with a brief comparison with the optimized k-Nearest Neighbours and Support Vector Machine for benchmarking. Section IV provides a discussion of the results followed by the suggestions for future work.

## II. INSTANCE-BASED TEMPLATE REGENERATION

The Instance-based Adaptive Template Reconstruction (I-ATR) algorithm proposed here consists of two phases (cf. Figure 1). Phase 1 (Training phase) generates intermediate templates by reconstructing/transforming the initial templates, so that the between-class separation is enhanced. Phase 2 (Test phase) further transforms these intermediate templates based on the query patterns, whilst transforming the query patterns in the meantime, to establish the best class assignment for the given query. This is accomplished by separately reconstructing the intermediate templates for each class, in turn, so that the distance of the query from the nearest class template is the smallest achievable. Each query may be composed of several feature vectors captured from segments of the data recording under observation. This set of instances would then be amalgamated into a template before being assigned to one of the possible classes. The reconstruction of the query is determined by comparing the shortest distances between the reconstructed templates and the transformed query for all the candidate classes in turn. The classification decision would then be determined for the transformed patterns showing the best match.

The following describes the proposed algorithm in more detail. Consider an $N$-class classification problem. $T$ denotes the instances of the training set for this problem, the query (test) instances are denoted as $Q$, and the instances per class as $I(n)$, where $n$ indicates the class label ($1 \leq n \leq N$). Each instance is a $L$- dimensional feature vector, while each class may contain a different number of instances. A three-dimensional matrix $T_{n,i,l}$, is used to represent the data elements of the feature vector in the training set. Here $i$ denotes the index of an instance, $n$ denotes the class to which that instance belongs, and $l$ denotes the feature dimension index ($1 \leq l \leq L$) (cf. Figure 2).

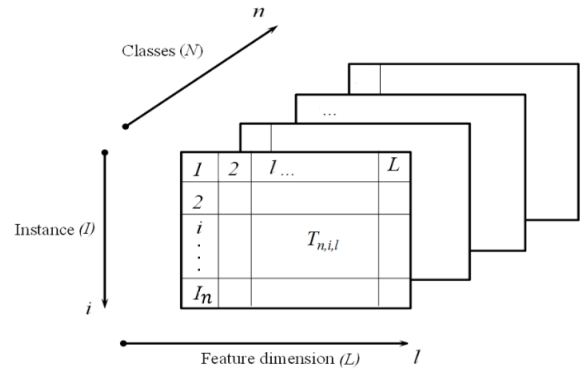

Fig. 2. Training set matrix $T_{n,i,l}$, where $1 \leq n \leq N$, $1 \leq i \leq I(n)$, $1 \leq l \leq L$.

$Q_{r,l}$ , $(1 \leq r \leq R)$, represented the query data where $R$ is the number of feature vectors extracted from one data recording. In a typical pattern classification task, $R = 1$; but for time-series data (such as EEG), features can be extracted from windows/epochs of the same recording; thus, resulting in multiple feature vectors from one recording ($R > 1$).

Phase 1 of the proposed I-ATR algorithm is devoted to maximising the between-class separation of the training set. Within each class, $n$, and for each feature element, $l$, of a training instance (feature vector), $i$, within that class, the average distances, $\bar{d}_{n,i,l}$, between it and the corresponding feature elements of the training feature vectors of the other classes are exhaustively measured, i.e., step 1) - step 2) in Block 1. These mean distances $\bar{d}_{n,i,l}$ for a given class $m$ $(1 \leq m \leq N)$, indicate the distance scores for the corresponding feature elements with respect to other classes. The elements are then ranked based on their respective scores and the values of a subset of $K$ the elements with large scores are retained to form the intermediate templates, $T'_{n,i,l}$, shown by step 3) - step 4) in Block 1. Depending on the application, $K$ could be empirically fixed. The full steps of Phase 1 of the I-ATR algorithm are shown in Block 1.

In the proposed method the distances between elements of each dimension of the feature vectors from different classes are measured, instead of the distance between the feature vectors. It may be worth noting that the Phase 1 of the algorithm has some similarities with the computation of Hausdorff Distance [14]. The reconstructed instances in $T'_{n,i,l}$ (here, $1 \leq i \leq K$) are created to have better between-class separation. As different feature vectors from the original set $T_{n,i,l}$ are used to reconstruct/transform these feature instances, the intermediate templates in $T'_{n,i,l}$ may not have exact equivalents in the original set of feature vectors. However, these reconstructed vectors will always be on the surfaces of, or within the bounding hyper-plane of the original training data set.

Phase 2 of the proposed method reconstructs the intermediate templates further, in conjunction with the query vectors, to produce $T''_{n,i,l}$ to minimise the distances between $T''_{n,i,l}$ and the transformed query feature vectors $Q_{r,l}$. When multiple instances from the same query observation are available (such as EEG data), the adaptive reconstruction is also applied to the instances from the query set, producing transformed query vectors $Q'_{r,l}$. As will be demonstrated in Section 3, this phase appears to be effective in alleviating the template longitudinal variations in person recognition using EEG signals.

As in Phase 1, distances $d_{n,i,l}$ between the intermediate template $T'_{n,i,l}$ for each candidate class and the query patterns $Q_{r,l}$ are measured in Step 5), Block 2. However, unlike Phase 1, only the feature elements contributing the smallest a few distances are retained to form the newly transformed feature vectors $T''_{n,l}$ and $Q'_{s,l}$, the process is illustrated by the Step 6) – Step 9) in Block 2. The distance between $T''_{n,l}$ and $Q'_{s,l}$ represents the score for the class under consideration. The process is repeated for each class, thus ensuring the best possible match between the training and the test data for each class without bias. The label of the class achieving the lowest distance score(s) (hence, most similar) is assigned to the query

---

**Phase 1: Maximize between-class Separation**

For $n = [1, N]$
   For $i = [1, I(n)]$
      For $l = [1, L]$
         1) Compute $d_{n,i,l}(m,j) = distance(T_{n,i,l}, T_{m,j,l})$,
         where $1 \leq m \leq N, m \neq n$ and $1 \leq j \leq I(m)$.
         // any appropriate distance metric may be used
         here. In this work, *distance* $(x, y) = |x\text{-}y|$ is used.
         2) Compute the mean of the resulting distances:  :
         $\bar{d}_{n,i,l}(m) = \frac{1}{I(m)} \sum_j d_{n,i,l}(m,j)$.

End of for-loops.
For $n = [1, N]$
   For $l = [1, L]$
      3) $index = arg\ sort_i(\bar{d}_{n,i,l}, descending)$.
      4) $T'_{n,i,l} = T_{n,index(1:K),l}$, where $K$ is the number of
      intermediate templates to be generated for each
      class.
End of for-loops.

// The resulting feature set $T'_{N,I,L}$, where $I = index(1:K)$, $K < I(n)$, now forms intermediate templates with relatively large between-class distances, which will be used in Phase 2 of the process.

Block 1. Phase 1 of the proposed I-ATR algorithm

---

**Phase 2: Preferential matching between the query sample and the stored templates**

// Let $T'_{n,i,l}$ are the intermediate templates (from Phase 1) and $Q_{r,l}$ be the query feature vectors from one observed sample. $T'_{n,i,l}$ has $K$ samples per class generated in Phase 1 and $Q_{r,l}$ has $R$ instances extracted from an observation.

For $n = [1, N]$
   For $i = [1, K]$
      For $l = [1, L]$
         5) Compute $d_{n,i,l}(r) = distance(T'_{n,i,l}, Q_{r,l})$,
         where $1 \leq r \leq R$.
End of for-loops.
For $n = [1, N]$
   For $l = [1, L]$,
      6) $index = arg\ sort_i(d_{n,i,l}, ascending)$.
      7) $T''_{n,l} = T_{n,index(1:F),l}$, where $F$ is the number of finally
      preserved instances for the training set per class.
      // Define the instance index for query set as index_q:
      8) $index\_q = arg\ sort_r(d_{n,i,l}(r), ascending)$.
      9) $Q'_{s,l}(n) = Q_{index_{q(1:S)},l}$ , where $S$ is number of
      modified query vectors to be reconstructed.
End of for-loops   //

// Classification Stage
For $s = [1, SN]$
   10) Compute $\tilde{n}(s) = \arg min_s\ distance(T''_{n,l}, Q'_{s,l})$
// $\tilde{n}(s)$ is the class closest to the query under consideration
End of for-loop
$overall\ decision = majorityVote(\tilde{n}(s))$.

Block 2. Phase 2 of the proposed I-ATR algorithm

pattern, as is shown by Step 10). The full algorithmic process for Phase 2 is illustrated in Block 2.

A few features can be distilled from the proposed I-ATR algorithm: it has been pointed in [15], if enough knowledge is available about what make the samples similar/dissimilar in the problem domain, the instance-based methods could be a good choice. In this context, the original samples are purposely reconstructed based on explicitly defined rules (Block 1&2) in the proposed work. This is the motive of performing template reconstruction within the instance-based learning framework.

In addition, it is observed that as the data volume and dimensionality increasing, the mass of a multivariate Gaussian distributions are pushed toward to the distant surface of the hypersphere, which makes many similarity measurements no longer effective [15]. It is proposed in this work to directly construct / generate new templates from the existing data pool, to bypass this predicament with expanded feature extensions in the hyperspace. Studies have also indicated that the rectilinear distance metric ($L_1$ norm) is consistently more preferable than distances with higher norms, such as the Euclidean distance metric ($L_2$ norm) for high dimensional data mining applications [16]. Therefore, without violating the Triangle Inequality, the $L_1$ norm distance measurement between elements from different instances (or feature vectors) is used in the I-ATR algorithm to construct new templates.

A hyper bounding box is established in the Cartesian space subsequently. Within this hyper bounding box: each dimension / attribute from the newly constructed templates is to exhibit the lowest between-class similarities possible, i.e., features distribute around the upper boundary of the hypersurface (after Phase 1); as well as the shortest distances possible between the reconstructed query and template feature vectors for all the training classes, i.e., features distribute around the closest boundaries between the training and query hypersurfaces (in Phase 2). It also worth to point out, though the distance of the reconstructed vectors through the proposed mechanism were directly measured in this study, they may also be effective for other classification algorithms, such as neural nets and linear classifiers.

## III. CASE STUDY EVALUATIONS

In this section, the proposed method is evaluated using EEG data to explore its robustness across different application scenarios. Two major subsections are subsequently presented, each devoted to evaluating the I-ATR classification method using two EEG databases respectively.

Evaluation studies are presented to assess the effectiveness of the proposed method for the classification of 1-D time series data. Unlike the 2-D images, the EEG signal in particular, due to its nonstationary nature, could be more challenging for successful pattern recognition. In this section, two EEG databases, one publicly available, the other self-collected are used to further evaluate the proposed method.

Section 3.1 reports the evaluations based on a publicly available database collected using BCI2000 instrumentation system [17], namely the EEG Motor Movement/Imagery Dataset (EEG MM/I) [18]. The results for biometric person recognition from 105 selected classes (subjects) from the MM/I database are presented.

In Section 3.2, a more challenging dataset namely Mobile Sensor Database is used to investigate the efficacy of the proposed algorithm for person recognition, where the signal quality is low and only a relatively small amount of training data is available. The impact of the template ageing effect due to multiple sessions with substantial time intervals is also addressed in the last set of the experiments.

### 3.1 MM/I Dataset for Evaluation of Biometric Recognition

This section explores the effectiveness of the I-ATR algorithm in person recognition problem, the results are compared with optimised 1-NN and SVM algorithms. To ensure at least 2 minutes of recording is available for each subject, 105 subjects (classes) were selected from the MM/I dataset. The studies reported in this section used the wavelet-based features were described in [19].

In the work reported here, to simulate application scenarios where ease of deployment and fast data processing is important, only the data from the Oz electrode of the MM/I dataset is used purposely. The experimental arrangement is briefly described below:

1) The EEG signals of four mental tasks (T1-T4) were segmented into multiple 4-second overlapping epochs (50% overlapping).
2) The wavelet packet decomposition (WPD) [20] was carried out for each time-domain window up to level-3. The resulting wavelet coefficients from level-3 between 0 and 60 Hz were retained for feature extraction. Every generated feature vector had six dimensions (each dimension approximately corresponding to a bandwidth of 10 Hz).
3) The variance of the wavelet coefficients in each epoch was used as a feature.
4) The I-ATR algorithm was then invoked for the process of template generation and classification.

### 3.1.1 Parameter Optimization

Prior to the invocation of the I-ATR algorithm, the EEG recordings were segmented into 240 windows ($I = 240$). Note, the training set contained two two-minute recordings for each run, each accounted for 120 windows. The highest recognition rate was achieved while the parameter $K = 160$, $K$ indicates the amount of reconstructed and preserved feature templates for the Phase-1 training. Here the optimal proportion of the reconstructed instances to retain is empirically established, particularly for person recognition applications using the EEG MM/I dataset.

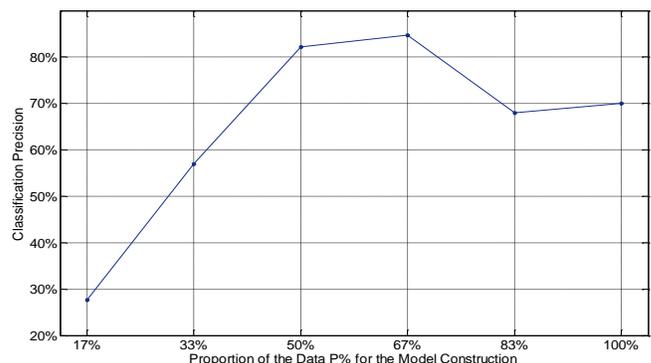

Fig. 3. Preliminary tests for optimizing $P\%$, the analyze was done by dividing the training set into six roughly equal size portions, the preserved data for training is incrementally tested.

Recognition accuracies were calculated by the leave-one-out three-fold cross-validation approach using in total 6 minutes of the recording from each task. The volume of the retained newly constructed feature vectors is defined as *P%* (of the available total) which corresponds to the absolute parameter *K* in Step 4 of Phase 1 shown in Block 1. The recognition results for various *P* are illustrated in Figure 3.

Figure 3 shows the change in classification accuracy while *P%* of the intermediate templates were preserved in the Phase 1 of the I-ATR algorithm, these intermediate templates were reconstructed from the initial instances after feature extraction (240 windows/observations). As a preliminary exploration for parameter optimisation, here only a small subset of the dataset (S1 - S10) was used to investigate the overall impact of the retained volume of the reconstructed feature instances for the follow-up Phase 2 processing. This is also consistent with some real-world training constraints where availability of data would be limited. The impact of *P%* on classification performance was investigated where only the Phase 1 of the proposed algorithm was applied (as depicted on the horizontal axis in Figure 3).

Results indicate that higher performances are achieved when the retained volume of reconstructed intermediate templates is about half to 2/3rd of the available training data. Therefore, as a rule of thumb, roughly 2/3 of the reconstructed templates will be retained for the next stage of the algorithm. This parameter should be dependent on the quality of the database and the application scenario: served as a data quality indicator, for low-quality and noisy signals such as the Mobile Sensor Database, a different value of *P* may be optimum.

### 3.1.2 Person Identification using MM/I Dataset

Following the empirical optimisation of parameter *P*, only about 2/3 of the training subset of the MM/I dataset samples in their reconstructed form are retained for subsequent evaluation of the proposed method in Phase 2. Comparisons are also made with the classical k-NN (k = 1) and nonlinear SVM [21], as both of them are popular instance-based classification algorithms. Figure 4 presents the resulting cumulative match characteristic (CMC) [22] curves for person recognition using EEG MM/I dataset. The I-ATR algorithm is seen to result in the best top-1 identification rate of 92.6%. The 1-NN classifier produced the higher recognition rates only after rank-13. Both the SVM and 1-NN had less than 50% accuracy. It is worth to mention that we did not conduct performance comparisons with deep neural nets in this case study for two main reasons: 1) the available data per class for the EEG MM/I dataset is not large enough for the training to converge for this classification problem, 2) the Mobile Sensor Database used for comparison is an even smaller and more challenging cohort.

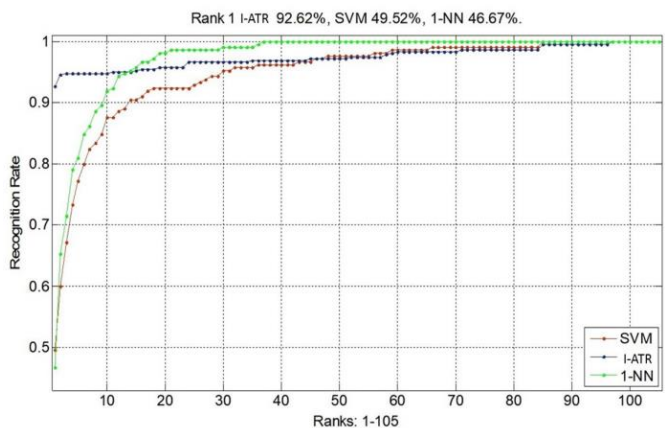

Fig. 4. Average recognition performances with three algorithms, using the MM/I database with 105 subjects. The SVM with 2nd order polynomial kernel and 1-NN are optimized for this scenario.

### 3.1.3 Person Verification using MM/I Dataset

Figure 5 presents the performance comparison of the three algorithms using the detection error trade-off (DET) characteristics curves [23]. From the perspective of biometric person verification, Phase 1 and Phase 2 of the I-ATR algorithm can be viewed as a mechanism to reduce the false acceptance rate (Phase 1, reducing the between-class overlaps in the training set) and false rejection rate (Phase 2, enhancing the provisional within-class matching between training classes and the query), respectively.

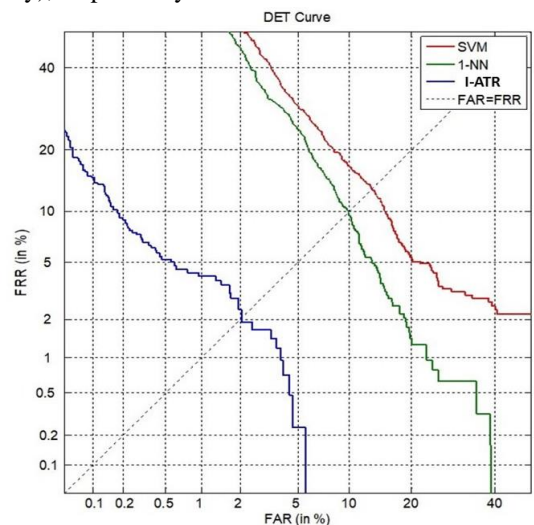

Fig. 5 DET curves of three learning algorithms, using the MM/I database with 105 subjects.

For the I-ATR algorithm, each feature vector was generated/reconstructed from 4-second recordings of the EEG signal, which producing 640 samples at 160 Hz. However, for the SVM and 1-NN algorithms, 6400 samples per window was used to maintain their best performances. Figure 5 shows an approximately 2% averaged equal error rate (EER) for the I-ATR algorithm whereas for the 1-NN and SVM classifiers the averaged EER values were approximately 10% and 14% respectively.

### 3.2 Performance using Mobile Sensor Database

The person recognition performance of the I-ATR algorithm was also tested using the locally collected Mobile Sensor Database to show the impact of a more challenging scenario. This database was purposely collected to explore the EEG template ageing effect in biometric scenarios. The experimental scheme was aimed to mimic the real-life practice: with low-cost single dry sensor and relatively long time-interval (three to eight weeks) between the sessions. The EEG data recorded from 30 participants (classes) were used for this study. For the multi-session evaluation, 4 seconds of continuous data chunks were randomly selected from each 1-minute recording of one session to form a set of the query attempts, and all the data from the other session was used for training the system. The key features of this database are as follows:

1) Data was captured by a gaming-grade single (Fp1 positioning electrode) dry sensor system (NeuroSky MindWave [24]), designed for ease of deployment.

2) EEG records were collected from 30 individuals (age ranges from 21 to 55) without any neurological background.

3) Subjects were required to engage in a simple sub-vocal number-counting activity (with eyes closed), i.e., the subjects sit in a silent room counting numbers from 1 onwards, the EEG data were recorded in the meantime.

4) Data was collected in two sessions with the time interval between the sessions ranging from three to eight weeks.

A well-established feature, based on wavelet transform was employed for the analyses presented in this study [19].

### 3.2.1 Longitudinal Template Ageing Effect

The I-ATR algorithm was in part motivated by the challenge of the template ageing effect in EEG biometric recognition systems, especially where only limited data is available for training [3]. In this section, the impact of using EEG data from different sessions with respect to different classification algorithms are presented. This is highlighted by evaluating the effectiveness of different I-ATR phases in dealing with such template variation phenomenon.

Table 1 shows the impact of template ageing when using EEG recordings from the Mobile Sensor Database for person identification. The same parameters as in the last section were used for the 1-NN and SVM classifiers, and their best performances for this database are reported for comparison. For the single-session scenario, the 60 seconds' recording was split into five non-overlapping segments to enable cross-validation. The average accuracies when using SVM and 1-NN classifiers have been found to be comparable at around 93%.

In the case of multi-session data analysis, the entire Session 1 data (60 seconds) was used for training. A randomly selected segment of the data from Session 2 was then used for testing. The training-testing order were then swapped, and the classification results averaged. The impact of template ageing is found to be quite significant, as is shown in Table 1. Only around 10% recognition accuracy were achieved with the conventional instance-based classifiers.

The impact of the two phases with respect to the proposed I-ATR algorithm has also been noticed. As it is illustrated in Table 1, with only the Phase 1, the classification performance in the double-session case (for template ageing) shown substantial improvement compared to 1-NN and SVM with 2nd order polynomial kernel. This indicates the great effectiveness of constructing new feature vectors with better between-class separation. By only applying Phase 2, similar considerable improvement was also observed; however, due to the absence of Phase 1 to reduce the potential between-class similarity, the false acceptances were not effectively limited, as the feature instances between the classes in the training set were not well-separated. It is clear from the results that when both the phases are employed, performance revealed drastic improvement, in particular for the multi-session scenario.

### 3.2.2 Mobile Sensor Database in Identification Scenario

For the biometric identification scenario, the CMC results of the I-ATR algorithm are shown in Figure 6. The recognition results obtained using the conventional classifiers are also included for comparison.

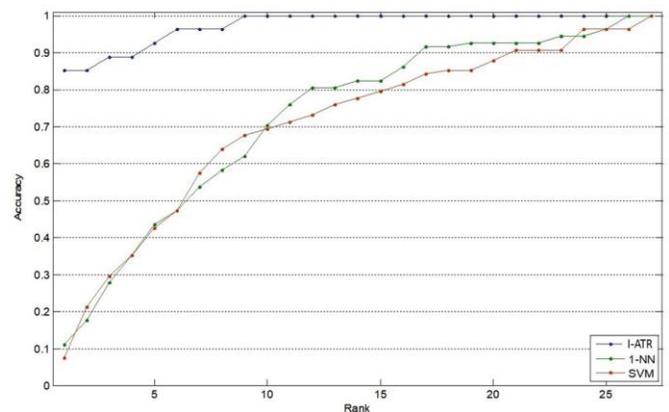

Fig. 6 CMC curves of the three learning algorithms, using the Mobile Sensor database, captured from 30 subjects using single EEG sensor at Fp1 location.

The results for these plots were generated using the Session 1 EEG data (60 sec. recording) for classifier modelling and 4 sec. recordings from Session 2 as query sets. The initial parameter $I = 240$, i.e., the total number of feature instances before template reconstructions. The finally retained number of instances in Phase 2 is $F = 4$ for both the training and query sets. As it can be clearly observed in Figure 6, the I-ATR method provided much higher rank-1 accuracy (85.71%) than the conventional classifiers.

### 3.2.3 Mobile Sensor Database in Verification Scenario

The DET curves for the Mobile Sensor Dataset are presented in Figure 7. Due to the challenging experimental setup (limited and noisy EEG data from a low-cost single dry sensor, as well as cross-session test data), the EER values of all the illustrated algorithms were found to be not as low as those with the MM/I dataset (Figure 5). However, it is clear that the I-ATR algorithm still produced much superior EER than the other two conventional instance-based classifiers. The EER of I-ATR method is 7.5%, whereas both the 1-NN and SVM schemes' EER values are higher than 30%.

TABLE 1 Session impact in EEG biometric identification, the comparisons are made among 1-NN, SVM and I-ATR algorithms. Using the Mobile Sensor Database.

| Template Stability | Mean Classification Accuracy (%) | | | | |
|---|---|---|---|---|---|
| | 1-NN | SVM | I-ATR Phase 1 | I-ATR Phase 2 | Full I-ATR |
| Single Session | 93.61 | 93.24 | 95.35 | 96.54 | 98.76 |
| Multi Session | 11.23 | 10.10 | 53.57 | 59.29 | 85.71 |

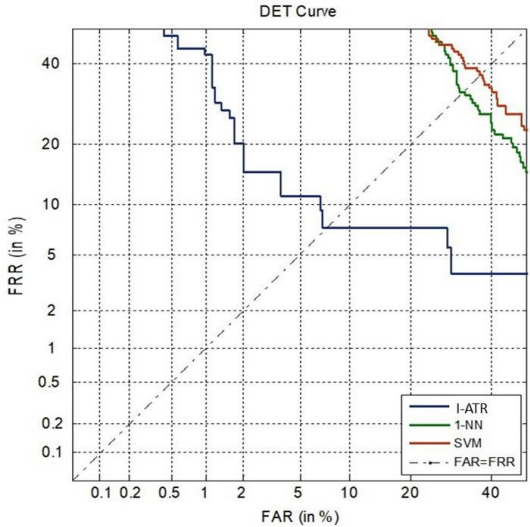

Fig. 7 DET curves of three learning algorithms, using the Mobile Sensor database with 30 subjects.

The results in Figure 5 and Figure 7 also highlight the difference in the data quality between the EEG MM/I dataset and the Mobile Sensor Database. The EER using I-ATR for the EEG MM/I database was as low as 2% whereas for the Mobile Sensor Database with only 30 subjects, it was about 7%. These analyses indicate the Mobile Sensor Database is indeed a much more challenging database for algorithm evaluation.

## IV. DISCUSSION AND CONCLUSION

In this paper, a novel instance-based template reconstruction classification algorithm, namely the I-ATR recognition method was presented, and its underlying motivation as well as rationale explained. Extensive comparative evaluations using time series data (EEG) from different resources demonstrated the effectiveness and robustness of the proposed method. The I-ATR classification algorithm shows superior performance and surpassed the currently reported results for the respective databases in the explored application scenarios. Specifically, the performance of the I-ATR algorithm was compared with two other popular and closely related instance-based classification algorithms, k-NN and SVM, using two EEG databases with considerably different characteristics. The methods based on deep neural nets were purposely excluded for reporting in this study, as the model training is unable to successfully converge due to the rather limited amount of data volume for each class.

The proposed algorithm was also shown to have resulted in substantial performance improvements in different recognition scenarios. In particular, the resilience of the I-ATR to template ageing even with input samples of limited recording length (2 to 6 minutes) was notable. It is thanks to the proposed approach potentially generates a substantially larger volume of reconstructed data from the limited available data, and subsequently a small subset with reconstructed features is retained to ensure the superior class separation. The intermediate templates (Phase 1 and Phase 2) help to overcome the challenge of limited data.

Compared to conventional classification approaches, the major difference of the proposed method lies in the element-wise template reconstruction for improved class separation, followed by the decision-making based on reconstructed feature vectors. In this approach, each feature dimension is treated independently and thus the space of all possible feature vectors that could be reconstructed is vastly increased. It is a fundamental assumption of this approach that this expanded space is nevertheless broadly representative of the underlying distribution for each class. Furthermore, it is assumed that the resulting feature reconstructions based on this expanded class representations result in new feature vectors that can be better separated using the vector-wise Euclidian distance calculations adopted for classification.

The proposed approach shares certain similarities with some recent pattern recognition algorithms in its motivation and conceptual principles. In the image processing field, a number of algorithms have recently been proposed to deal with the issue of pedestrian image alignment [25]. The alignment between the query image and the image in the database is one of the major challenges due to the variant camera angles and pedestrian postures. The pose attention transform [26] and the alignment affine transform [27] were recently developed to construct new images based on the query, which provided better matches in the database to achieve improved identification performances. The proposed I-ATR method also confines the reconstruction within the "bounding box" region [27] in order to guarantee the outcome of the feature transformations does not lose the ability to represent the original signal source.

At the same time, in contrast to the approaches implemented in the image/video processing field, the proposed template reconstruction method is motivated by some special requirements of bio-signals (such as EEG). The size of many publicly annotated EEG data is much smaller compared to the vast quantity of image data available for machine learning. Therefore, as opposed to the usage of deep nets in [26], [27], the I-ATR method is designed to leverage a relatively small amount of data, from which to produce high-quality features in order to achieve improved recognition performance.

Another distinguishing characteristic of the I-ATR algorithm is that it reconstructs the feature vectors of both the training set and the query for each recognition event, thus allowing improved flexibility in controlling the system performance (in terms of sensitivity vs. specificity), as required for specific application scenarios. The proposed approach has, therefore, the potential for applications in other domains using different modalities.

One focus of further work will be to establish the wider applicability of the proposed approach using other databases, including those from domains other than time-series data. Also, attention will be paid to the optimization of the algorithm through the adjustment of its parameters and its sensitivity to outliers. Deep learning methods may be incorporated with the proposed I-ATR algorithm to further improve the performance of the proposed approach.

### REFERENCES

[1]    J. P. Li Deng, L. Deng, and J. Platt, "Ensemble Deep Learning for

Speech Recognition," *Proc. Interspeech*. 2014. [Online]. Available: http://193.6.4.39/~czap/letoltes/IS14/IS2014/PDF/AUTHOR/IS140 245.PDF%5Cnhttps://www.microsoft.com/en-us/research/publication/ensemble-deep-learning-for-speech-recognition/.

[2] Mesnil, G., Dauphin, Y., Yao, K., Bengio, Y., Deng, L., Hakkani-Tur, D., He, X., Heck, L., Tur, G., Yu, D. and Zweig, G., "Using Recurrent Neural Networks for Slot Filling in Spoken Language Understanding," *IEEE/ACM Trans. Audio, Speech, Lang. Process.*, vol. 23, no. 3, pp. 530–539, 2015, doi: 10.1109/TASLP.2014.2383614.

[3] S. Yang and F. Deravi, "On the Usability of Electroencephalographic Signals for Biometric Recognition: A Survey," *IEEE Trans. Human-Machine Syst.*, vol. 47, no. 6, pp. 958–969, 2017, doi: 10.1109/THMS.2017.2682115.

[4] I. Hendrickx and A. Den Van Bosch, "Hybrid algorithms with instance-based classification," *Lect. Notes Comput. Sci. (including Subser. Lect. Notes Artif. Intell. Lect. Notes Bioinformatics)*, vol. 3720 LNAI, pp. 158–169, 2005, doi: 10.1007/11564096_19.

[5] W. Daelemans and A. Van den Bosch, "Memory-based language processing," *Cambridge Univ. Press*, no. January, pp. 1–189, 2005, doi: 10.1017/CBO9780511486579.

[6] D. Wilson and T. Martinez, "Reduction techniques for instance-based learning algorithms," *Mach. Learn.*, pp. 257–286, 2000, Accessed: Sep., 2015. [Online]. Available: http://link.springer.com/article/10.1023/A:1007626913721.

[7] I. Triguero, J. Derrac, S. García, and F. Herrera, "A taxonomy and experimental study on prototype generation for nearest neighbor classification," *IEEE Trans. Syst. Man Cybern. Part C Appl. Rev.*, vol. 42, no. 1, pp. 86–100, 2012, doi: 10.1109/TSMCC.2010.2103939.

[8] P. Hart, "The condensed nearest neighbor rule (Corresp.)," *IEEE Trans. Inf. theory*, vol. 14, no. 3, pp. 515–516, 1968.

[9] J. A. Olvera-López, J. A. Carrasco-Ochoa, J. F. Martínez-Trinidad, and J. Kittler, "A review of instance selection methods," *Artif. Intell. Rev.*, vol. 34, no. 2, pp. 133–143, 2010, doi: 10.1007/s10462-010-9165-y.

[10] J. A. Olvera-López, J. A. Carrasco-Ochoa, J. F. Martínez-Trinidad, and J. Kittler, "A review of instance selection methods," *Artif. Intell. Rev.*, vol. 34, no. 2, pp. 133–143, May 2010, doi: 10.1007/s10462-010-9165-y.

[11] S. García, J. Derrac, J. R. Cano, and F. Herrera, "Prototype selection for nearest neighbor classification: Taxonomy and empirical study," *IEEE Trans. Pattern Anal. Mach. Intell.*, vol. 34, no. 3, pp. 417–435, 2012, doi: 10.1109/TPAMI.2011.142.

[12] M. P. Tarvainen, J. K. Hiltunen, P. O. Ranta-aho, and P. A. Karjalainen, "Estimation of nonstationary EEG with Kalman smoother approach: an application to event-related synchronization (ERS).," *IEEE Trans. Biomed. Eng.*, vol. 51, no. 3, pp. 516–24, Mar. 2004, doi: 10.1109/TBME.2003.821029.

[13] S. Theodoridis and K. Koutroumbas, *Pattern Recognition*, 4th ed. Elsevier, 2009.

[14] R. T. Rockafellar and R. J.-B. Wets, *Variational Analysis*. Springer Science & Business Media, 2009.

[15] D. Pedro, "A Few Useful Things to Know About Machine Learning," *Commun. ACM*, vol. 55, no. 10, pp. 78–87, 2012. [Online]. Available: https://dl.acm.org/citation.cfm?id=2347755.

[16] C. C. Aggarwal, A. Hinneburg, and D. A. Keim, "On the surprising behavior of distance metrics in high dimensional space," *Int. Conf. database theory*, pp. 420–434, 2001, doi: 10.1007/3-540-44503-x_27.

[17] G. Schalk, D. J. McFarland, T. Hinterberger, N. Birbaumer, and J. R. Wolpaw, "BCI2000: a general-purpose brain-computer interface (BCI) system.," *IEEE Trans. Biomed. Eng.*, vol. 51, no. 6, pp. 1034–43, Jun. 2004, doi: 10.1109/TBME.2004.827072.

[18] "EEG Motor Movement/Imagery Dataset v1.0.0." https://physionet.org/content/eegmmidb/1.0.0/ (accessed Dec., 2021).

[19] S. Yang and F. Deravi, "Wavelet-based EEG preprocessing for biometric applications," *Proc. - 2013 4th Int. Conf. Emerg. Secur. Technol. EST 2013*, pp. 43–46, Sep. 2013, doi:10.1109/EST.2013.14.

[20] I. Daubechies, *Ten lectures on wavelets*, Vol. 61. Philadelphia: Society for industrial and applied mathematics, 1992.

[21] D. M. J. Duin, R. P. W., Juszczak, P., de Ridder, D., Paclık, P., Pezkalska, E., & Tax, "PRTools," 2004.

http://37steps.com/prhtml/prtools.html (accessed Apr., 2015).

[22] D. Zhang and A. K. Jain, Eds., *Biometric Authentication*, vol. 3072. Berlin, Heidelberg: Springer Berlin Heidelberg, 2004.

[23] A. Martin, G. Doddington, T. Kamm, M. Ordowski, and M. Przybocki, "The DET Curve in Assessment of Detection Task Performance," *Natl. INST Stand. Technol. Gaithersbg. MD*, 1997, Accessed: Apr., 2015. [Online]. Available: http://oai.dtic.mil/oai/oai?verb=getRecord&metadataPrefix=html&identifier=ADA530509.

[24] "Neurosky Products." http://store.neurosky.com/products (accessed Nov., 2021).

[25] L. Zheng, Y. Yang, and A. G. Hauptmann, "Person Re-identification: Past, Present and Future," vol. 14, no. 8, pp. 1–20, 2016. [Online]. Available: http://arxiv.org/abs/1610.02984.

[26] Z. Zhu, T. Huang, B. Shi, M. Yu, B. Wang, and X. Bai, "Progressive pose attention transfer for person image generation," *Proc. IEEE Comput. Soc. Conf. Comput. Vis. Pattern Recognit.*, vol. 2019-June, pp. 2342–2351, 2019, doi: 10.1109/CVPR.2019.00245.

[27] Z. Zheng, L. Zheng, and Y. Yang, "Pedestrian alignment network for large-scale person re-identification," *IEEE Trans. Circuits Syst. Video Technol.*, vol. 29, no. 10, pp. 3037–3045, 2019, doi: 10.1109/TCSVT.2018.2873599.